\documentclass[aps,prl,twocolumn,superscriptaddress]{revtex4-1}

\usepackage[T1]{fontenc}
\usepackage[latin9]{inputenc}
\usepackage{amsmath}
\usepackage{amssymb}
\usepackage{graphicx}
\usepackage{hyperref}
\usepackage{natbib}
\usepackage{esint}
\usepackage{times}
\usepackage{helvet}
\usepackage{braket}
\usepackage{bbm}
\usepackage{xcolor}
\usepackage{enumitem}
 
\makeatletter
\newcommand{\JILA}{JILA and NIST, University of Colorado, Boulder, CO 80309, USA}
\newcommand{\PHYS}{Department of Physics, University of Colorado, Boulder, CO 80309, USA}
\newcommand{\CTQM}{Center for Theory of Quantum Matter, University of Colorado, Boulder, CO 80309, USA}

\begin{document}

\title{Quantum entropic self-localization with ultracold fermions}
\date{\today}

\author{M. Mamaev}
\email{mikhail.mamaev@colorado.edu}
\affiliation{\JILA}
\affiliation{\PHYS}
\affiliation{\CTQM}
\author{I. Kimchi}
\affiliation{\JILA}
\affiliation{\PHYS}
\affiliation{\CTQM}
\author{M. A. Perlin}
\affiliation{\JILA}
\affiliation{\PHYS}
\affiliation{\CTQM}
\author{R. M. Nandkishore}
\affiliation{\PHYS}
\affiliation{\CTQM}
\author{A. M. Rey}
\affiliation{\JILA}
\affiliation{\PHYS}
\affiliation{\CTQM}

\begin{abstract}
{We study a driven, spin-orbit coupled fermionic system in a lattice at the resonant regime where the drive frequency equals the Hubbard repulsion, for which non-trivial constrained dynamics emerge at fast timescales. An effective density-dependent tunneling model is derived, and examined in the sparse filling regime in 1D. The system exhibits entropic self-localization, where while even numbers of atoms propagate ballistically, odd numbers form localized bound states induced by an effective attraction from a higher configurational entropy. These phenomena occur in the strong coupling limit where interactions only impose a  constraint with no explicit Hamiltonian term. We show how the constrained dynamics lead to quantum few-body scars and map to an Anderson impurity model with an additional intriguing feature of non-reciprocal scattering. Connections to many-body scars and localization are also discussed.}
\end{abstract}
\maketitle

\textit{Introduction.}
Ultracold atomic systems loaded into optical lattices are among the most powerful quantum simulation platforms accessible today, especially when augmented with emerging capabilities for individual-particle manipulation. Dynamical control over internal and external degrees of freedom, lack of disorder, long coherence times and tunable interactions allow these systems to capture many of the ingredients at the heart of modern quantum science~\cite{bloch2012quantumSimulation}. These include gauge fields~\cite{osterloh2005seminal}, superconductivity~\cite{bloch2008seminal}, and even phenomena relevant to high-energy physics such as confinement~\cite{fradkin1979confinement}. However, optical lattice systems are often limited by slow effective cross-site interaction timescales such as superexchange in the Mott insulating regime.

In this work we look at a resonance-assisted model operating in the strongly interacting regime. This model gives rise to constrained physics where atomic motion is strongly correlated and species-dependent, while evolving at timescales set by the lattice tunneling rate. Our model employs a driving laser that interrogates internal pseudo-spin states, and imposes a relative phase between every lattice site which generates spin-orbit coupling (SOC)~\cite{wall2016synthetic, kolkowitz2016spinorbitcoupled,tai2017flux,bromley2018SOC}. The competition between the SOC-inducing drive and interactions makes one tunneling process resonant and inhibits the other processes, yielding an effective density-dependent tunneling. This setup is connected to prior experiments using tilted~\cite{aidelsburger2013tilted,meinert2014observation,miyake2013realizing,simon2011tilted,jurgensen2014tilted} and shaken lattices~\cite{lignier2007dynamical,meinert2016floquet,struck2011shaking,jotzu2014shaking,clark2018shaking,messer2018shakenLattices,gorg2018floquetTJModel} to study gauge fields~\cite{banerjee2012gaugeFields,bermudez2011gaugeFields,hauke2012gaugeFields,barbiero2018gaugeFields,gorg2018gaugeFields,schweizer2019gaugeFields}, transport~\cite{rubbo2011resonantly,ivanov2008acInducedTransport}, Mott-metal transitions~\cite{zenesini2009mottMetalShaking} and many other phenomena~\cite{eckardt2017seminalShaking}. Our work avoids the issue of heating effects caused by Floquet engineering, has a straightforward implementation, and explores subjects of scattering and localization that have not been considered as often in this context~\cite{preiss2015quantumWalks,lerose2019confinement}. This work is also related to the prior proposal of generating cluster states~\cite{mamaev2019clusterStates}, but now explores the non-perturbative regime where conventional superexhange breaks down.

The density-dependent model we derive exhibits multi-body self-localization, where particles localize themselves without additional external potentials. The effect is akin to a quantum entropic self-confinement, caused by the emergence of bound states induced by the minimization of kinetic energy in Fock states with higher connectivity. The bound states feature different physics than other mobile scattering states, a behavior that is reinforced by mapping the dynamics to that of an Anderson impurity model~\cite{anderson1961anderson,hewson1997anderson,kehrein1996anderson}, allowing us to derive analytic results on bound populations and scattering coefficients. This mapping is also supplemented by a particular type of asymmetric behaviour similar to those found in non-reciprocal systems~\cite{deak2012reciprocity}, manifesting as a scattering process where transmission moves a scatterer while reflection keeps it in place. These constrained dynamics are connected to the study of quantum scars~\cite{shiraishi2017scars,moudgalya2018scars,turner2018scars} as closed trajectories arise depending on the boundaries, drastically changing the long-time density profiles. 
Our results are also similar to those seen in Efimov physics~\cite{efimov1970efimov,naidon2017efimov,shi2014efimov}, since our system exhibits a three-body bound state and free-propagating two-body pairs. While the physics described here is at the level of a 1D system for which a hardcore boson description would yield equivalent results, the experimental implementation poses no significant restriction on dimension. This opens a wealth of prospects for further studies of even richer phenomena in higher dimensions, where Fermi statistics will matter and the features of quantum scars and non-reciprocity become qualitatively distinct.

The system can be realized experimentally in a number of platforms. The most promising one is 3D optical lattices using ultracold alkaline earth atoms, for which the long lifetimes and magnetic field insensitivity permit a clean study. Having faster timescales also permits implementation in systems with access to modern tools that allow for single-atom addressability, such as quantum gas microscopes~\cite{gericke2008quantumGasMicroscope,bakr2009quantumGasMicroscope,sherson2010quantumGasMicroscope,cheuk2015quantumGasMicroscope,yamamoto2016quantumGasMicroscope,mitra2018quantumGasMicroscope} and optical tweezers~\cite{endres2016tweezers,norcia2018tweezers,cooper2018tweezers,saskin2019tweezers}, which would facilitate easier observation of the dynamics.

\textit{Model.}
Our starting point is a cubic optical lattice of $L$ sites populated by $N$ fermionic atoms with two internal pseudo-spin states $g$, $e$ in the lowest Bloch band. The system is 3D in principle, although in this work we restrict to 1D by making the lattice confinement stronger along transverse directions. A resonant interrogating laser drives transitions between the internal states. The Hamiltonian is $\hat{H}=\hat{H}_{0}+\hat{H}_{\mathrm{\Omega}}$, where $\hat{H}_{0}=-J\sum_{\langle j,i\rangle,\sigma}(\hat{c}_{j,\sigma}^{\dagger}\hat{c}_{i,\sigma}+h.c.)+U\sum_{j}\hat{n}_{j,e}\hat{n}_{j,g}$ is the Fermi-Hubbard model (nearest-neighbour tunneling), and $\hat{H}_{\Omega}=\frac{\Omega}{2}\sum_{j}(-1)^{j}(\hat{c}_{j,e}^{\dagger}\hat{c}_{j,g}+h.c.)$ is the laser drive. Here, $\hat{c}_{j,\sigma}$ annihilates an atom of spin $\sigma \in \{g,e\}$ on site $j$, and $\hat{n}_{j,\sigma}=\hat{c}_{j,\sigma}^{\dagger}\hat{c}_{j,\sigma}$. The lattice tunneling strength is $J$, and the onsite repulsion is $U$. The laser drive has Rabi frequency $\Omega$ and a site-dependent phase $e^{ij\pi}=(-1)^{j}$ arising from a mismatch between the driving and confining laser wavelengths, generating spin-orbit coupling (SOC)~\cite{wall2016synthetic, kolkowitz2016spinorbitcoupled, tai2017flux}. Every adjacent pair of sites has a relative phase of $\pi$, resulting in a relative minus sign between their laser couplings.

\begin{figure}
\centering
\includegraphics[width=1\linewidth]{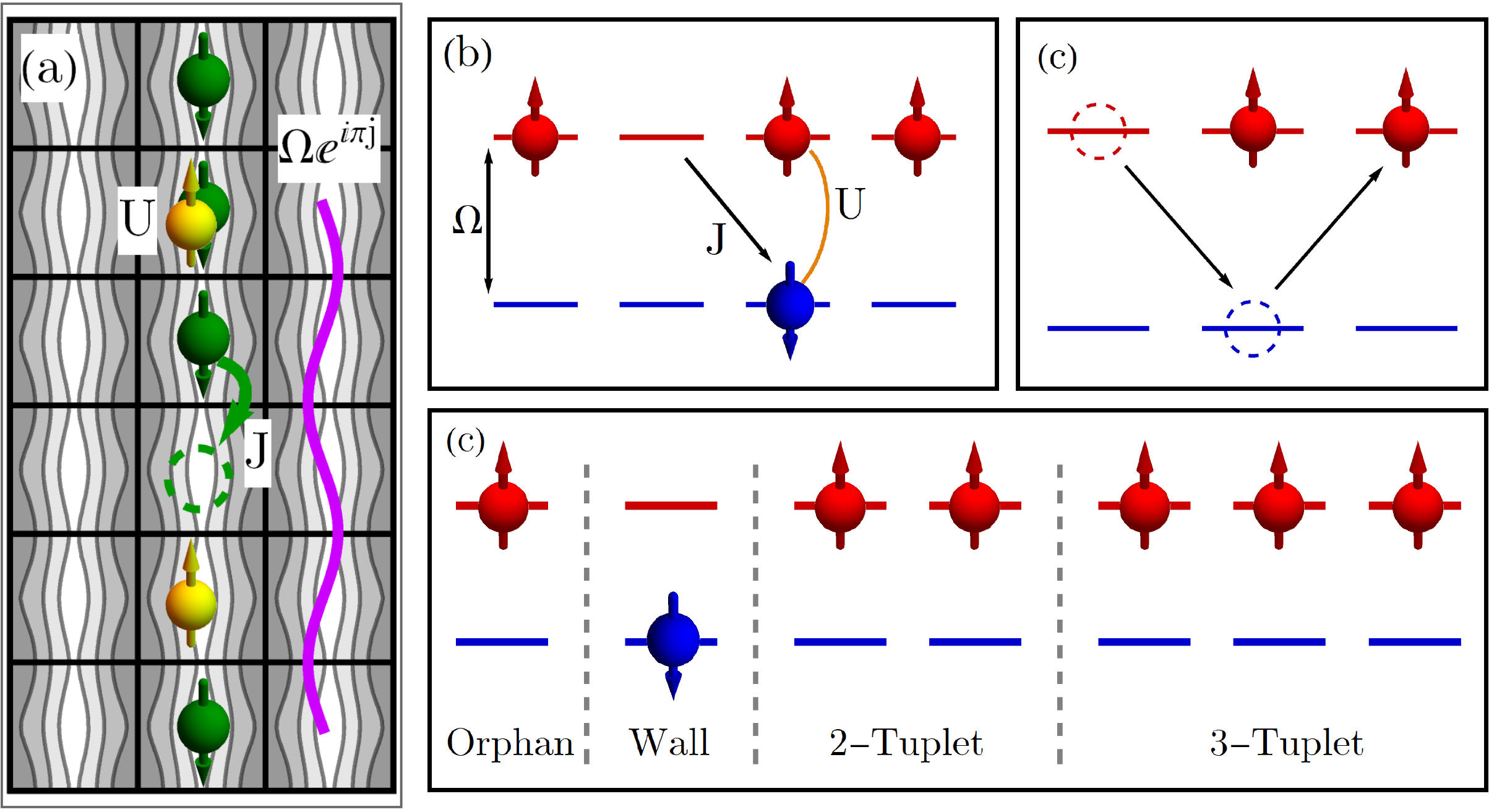}
\caption{(a) Schematic of Fermi-Hubbard optical lattice setup, confined to 1D, driven by an SOC-inducing laser with relative phase $\pi$ between sites. (b) Resonance-assisted atomic motion in the gauged frame. An $\uparrow$ atom can tunnel into a neighbouring site already holding an $\uparrow$ atom, flipping its spin and creating a doublon. All other tunneling processes are off-resonant. (c) Leap-frog motion of two $\uparrow$ atoms. (d) Few-body structures present in a sparsely filled lattice. Orphans cannot move on their own, but interact with other atoms. Walls inhibit all motion and cannot move. 2-tuplets move freely via the leap-frog mechanism. 3-tuplets can shoot off a 2-tuplet, but exhibit nontrivial three-body dynamics.}
\label{fig_Schematic}
\end{figure}

We define two new species of fermion, $\hat{a}_{j,\uparrow}=[\hat{c}_{j,e}+(-1)^{j}\hat{c}_{j,g}]/\sqrt{2}$ and $\hat{a}_{j,\downarrow}=[\hat{c}_{j,e}-(-1)^{j}\hat{c}_{j,g}]/\sqrt{2}$, for which the Hamiltonian becomes
\begin{equation}
\begin{aligned}
\label{eq_FermiHubbardGauged}
\hat{H}=&-J\sum_{\langle j,i\rangle}\left(\hat{a}_{j,\uparrow}^{\dagger}\hat{a}_{i,\downarrow}+\hat{a}_{j,\downarrow}^{\dagger}\hat{a}_{i,\uparrow}+h.c.\right)\\
&+U\sum_{j}\hat{n}_{j,\uparrow}\hat{n}_{j,\downarrow}+\frac{\Omega}{2}\sum_{j}\left(\hat{n}_{j,\uparrow}-\hat{n}_{j,\downarrow}\right),
\end{aligned}
\end{equation}
with number operators $\hat n_{j,\sigma'}=\hat a_{j,\sigma'}^\dag \hat a_{j,\sigma'}$ for new pseudo-spin states $\sigma'\in \{\uparrow,\downarrow\}$.
The laser drive in this frame acts as an effective magnetic field. Tunneling is accompanied by a spin-flip due to the SOC. Direct tunneling into empty sites must cross an energy gap $\pm\Omega$, and is inhibited for strong driving $\Omega/J \gg 1$.

However, if we tune the drive strength to match the repulsion $U=\Omega$, an $\uparrow$ atom can tunnel into a site already holding an $\uparrow$ atom by flipping its spin and creating a doublon (two atoms of opposite spin). The energy loss $-\Omega$ is resonantly compensated by an energy gain $U$ from repulsion, with total cost $-\Omega+U=0$ allowing the process to occur freely at rate $J$. A $\downarrow$ atom tunneling would instead cost $+\Omega+U = 2U$, and still be inhibited. Since tunneling into an already-occupied state is Pauli-blocked, the process described is the only one that can occur. In the limit $U/J=\Omega/J \to \infty$, we thus reduce the Hamiltonian to a density-dependent tunneling model (up to $J/U$ corrections),
\begin{equation}
\label{eq_FermiHubbardAsymptotic}
\hat{H}_{\infty} = -J \sum_{\langle j,i\rangle}\left[ \hat{n}_{i,\uparrow}(1-\hat{n}_{j,\downarrow})\left(\hat{a}_{i,\downarrow}^{\dagger} \hat{a}_{j,\uparrow}+h.c.\right)+(j \leftrightarrow i)\right],
\end{equation}
where the last term exchanges $j, i$, and the number terms ensure that the spin-flip tunneling has an $\uparrow$ atom on the destination and no $\downarrow$ atom on the origin. This model is valid for timescales $tJ \lesssim U/J$, see Appendix~\ref{app_ModelAgreement} for benchmarking.

The constrained motion is depicted in Fig.~\ref{fig_Schematic}(b), and can be understood as a leap-frog mechanism alternating between doublons and neighbouring $\uparrow$ singlons (single atoms). Assuming a sparse lattice with $N/L \ll 1$, we categorize the behaviour of few-atom configurations in Fig.~\ref{fig_Schematic}(c). An $\uparrow$ singlon is an orphan, which cannot move on its own but will interact with other atoms. A $\downarrow$ singlon can neither move nor permit motion through itself, acting as a wall. A 2-tuplet moves freely via the leap-frog mechanism. A 3-tuplet (three $\uparrow$ singlons) can also move by shooting off a 2-tuplet, but exhibits nontrivial three-body dynamics that we discuss below.

\begin{figure}
\centering
\includegraphics[width=0.9\linewidth]{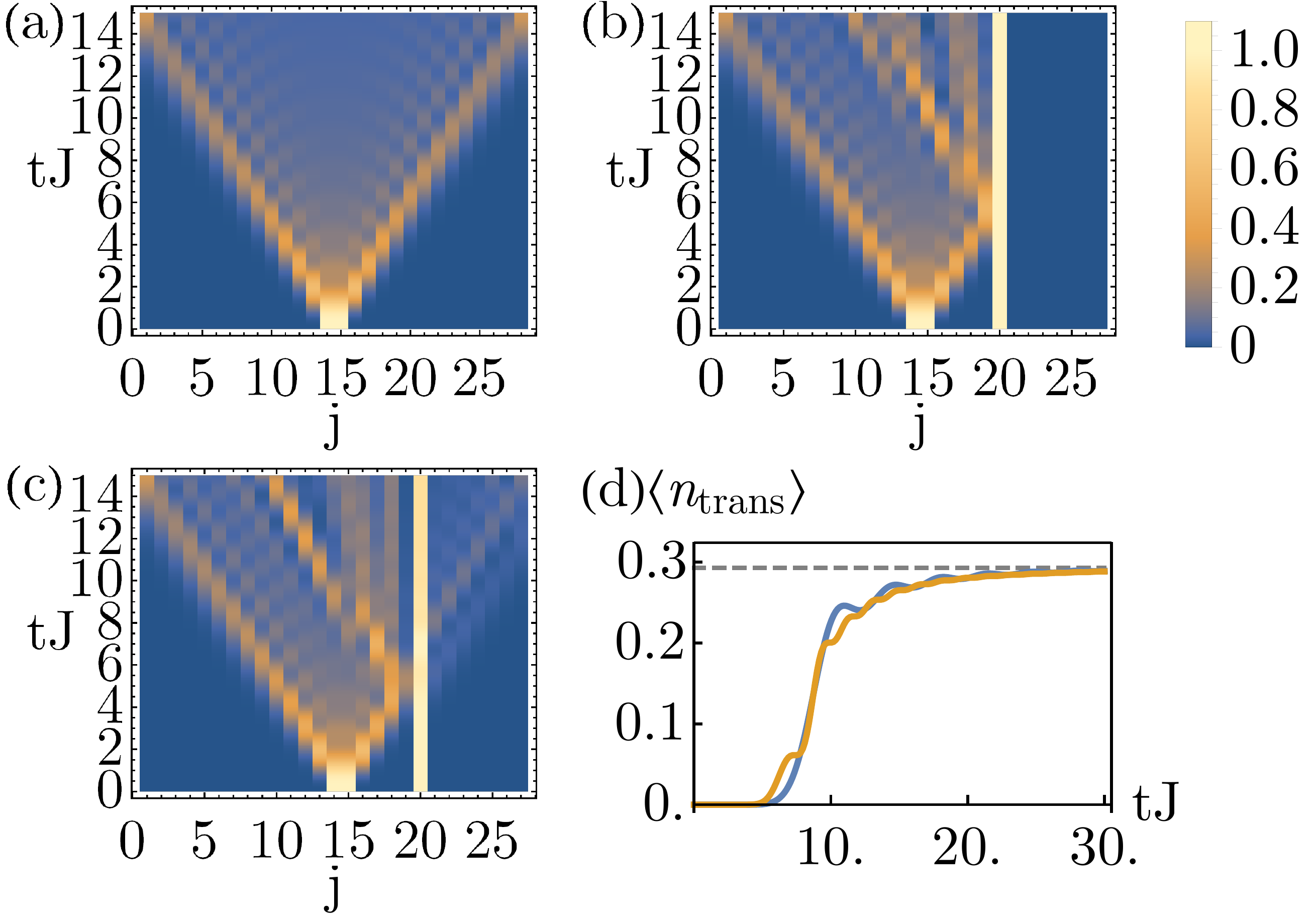}
\caption{(a) Dynamics of an initial 2-tuplet (using Eq.~\eqref{eq_FermiHubbardAsymptotic}). We plot atomic density $\langle \hat{n}_{j}\rangle$ (color scaled to emphasize features). (b) Reflection of a propagating 2-tuplet off a wall ($\downarrow$ singlon). (c) Collision of a propagating 2-tuplet with an orphan ($\uparrow$ singlon). Three-body dynamics cause non-reciprocal transmission and reflection. (d) Number of atoms transmitted past the orphan during the collision in panel (c). The orange line is direct Fermi-Hubbard numerics. The blue line uses a Kronecker delta potential single-particle scattering of strength $U_{\delta}=2J$, which has an asymptotic value of $\langle \hat{n}_{\mathrm{trans}}\rangle=1-1/\sqrt{2}$ indicated in gray (see discussion of scattering in text).}
\label{fig_DoubletMotion}
\end{figure}

In Fig.~\ref{fig_DoubletMotion}, we compare the density $\langle \hat{n}_{j}\rangle=\langle\hat{n}_{j,\uparrow}\rangle+\langle\hat{n}_{j,\downarrow}\rangle$ profiles of a 2-tuplet propagating freely (a), colliding with a wall (b), and with an orphan (c). The 2-tuplet acts as a quasiparticle that spreads ballistically, with wavefronts rising from an underlying quantum walk. Upon colliding with the orphan, we find that a nontrivial scattering process takes place. Some atom density $\langle \hat{n}_{\mathrm{trans}}\rangle=\sum_{j>j_0} \langle \hat{n}_{j}\rangle$ ($j_0$ the position of the orphan) is transmitted by forming a new 2-tuplet with the orphan, while some is reflected. The orphan acts as an effective scatterer whose own position is changed only if transmission occurs. The underlying reason for this outcome is tied to the self-localization properties of the three-atom system.

\begin{figure}
\centering
\includegraphics[width=1\linewidth]{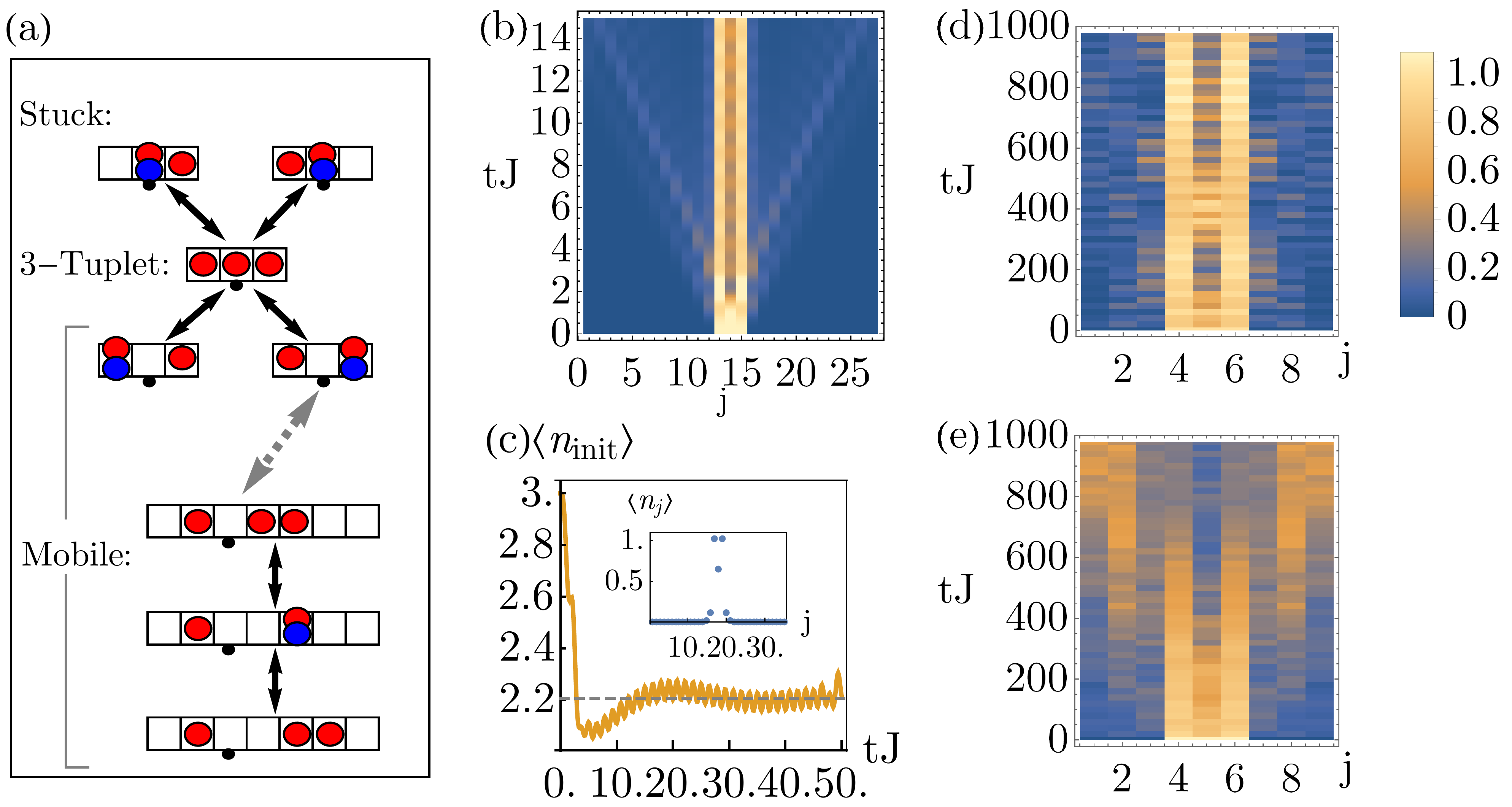}
\caption{(a) Schematic for a 3-tuplet's motion. Red and blue circles are $\uparrow$, $\downarrow$ atoms respectively. The 3-tuplet can tunnel into four states: Two stuck states, and two mobile (scattering) states that shoot off a 2-tuplet. The dots under the boxes indicate the center of the 3-tuplet. (b) Density dynamics for an initial 3-tuplet. (c) Number of atoms that stay in the initial three sites. The gray dashed line marks the analytic prediction of $\langle \hat{n}_{\mathrm{init}}^{(3)}\rangle\approx 2.2$ (c.f. Appendix~\ref{app_TripletBoundStates}). The inset shows the density profile for the 3-tuplet bound states $\ket{\phi_{\pm}^{(3)}}$ (equal for both). (d-e) 3-Tuplet density dynamics for long times $tJ \gg 1$ (coarse-grained in time), for a small lattice using open (d) and periodic (e) boundaries. The periodic boundaries eventually dissolve the localization by wrapping 2-tuplets around to pull the orphan away.}
\label{fig_Triplets}
\end{figure}

\textit{3-Tuplet bound states.}
The core dynamics for three atoms can be understood by considering a 3-tuplet. Fig.~\ref{fig_Triplets}(a) shows the states that it can tunnel into. An atom can either tunnel into the middle, yielding stuck states that cannot move further, or out from the middle to make a 2-tuplet free to propagate. The 3-tuplet state thus has connectivity $\nu =4$, which counts the number of Fock states that $\hat{H}$ couples it to. If we allow a 2-tuplet to propagate, an immobile orphan is generated, and the resulting state is reduced to connectivity $\nu = 2$.

This reduced connectivity lets us map the motion of the 3-tuplet to a 1D tight-binding chain with impurities. The accessible Fock states are left- and right-side 2-tuplet states, as shown in Fig.~\ref{fig_Triplets}(a). We associate each of these states with a virtual lattice coordinate $m$ (one for the doublon configuration, and one for neighbouring singlons), with two virtual sites $m$ for every real site $j$. The 3-tuplet acts as a central site coupling to the left- and right-chains of states. In addition, it also couples to the two stuck states, acting as extra dead-end links. This model is analogous to an non-interacting Anderson impurity model (c.f. Appendix~\ref{app_TripletBoundStates}),
\begin{equation}
\label{eq_AndersonModel}
    \hat{H}^{(3)}=-J\sum_{\langle m,n\rangle}\left(\hat{d}_{m}^{\dagger}\hat{d}_{n}+h.c.\right)-J\left(\hat{d}_{0}^{\dagger}\hat{d}_{L}+\hat{d}_{0}^{\dagger}\hat{d}_{R}+h.c.\right),
\end{equation}
where $\hat{d}_{m}$ annihilates a fermion on virtual site $m$, $\hat{d}_{0}$ on the 3-tuplet, and $\hat{d}_{L}$, $\hat{d}_{R}$ on the stuck states (acting as impurities). The Anderson impurity model is known to host bound states, and indeed, we observe their presence in our system as well. Fig.~\ref{fig_Triplets}(b) shows the density dynamics for an initial 3-tuplet. In contrast to the 2-tuplet, more than 2/3rds of the atoms remain in the sites they started in; see Fig.~\ref{fig_Triplets}(c), where $\langle \hat{n}_{\mathrm{init}}^{(N)}\rangle$ is the total atom number in the $N$ initially-filled sites. The 3-tuplet has overlap with exponentially localized bound eigenstates, and the corresponding population will not decrease as the system evolves. We find analytic expressions for the bound eigenstates, $\hat{H}_{\infty}\ket{\phi_{\pm}^{(3)}}=\pm E \ket{\phi_{\pm}^{(3)}}$ (up to exponentially small boundary corrections),
and for their energy $E = \sqrt{2}b J$ where $b^2=1+\sqrt{2}$ is the wavefunction localization length in units of the real lattice spacing (see Appendix~\ref{app_TripletBoundStates}).
In the limit $L \to \infty$, the 3-tuplet $\ket{\uparrow,\uparrow,\uparrow}$ has overlap $|\langle \uparrow,\uparrow,\uparrow|\phi_{\pm}^{(3)}\rangle|^2=1/2\sqrt{2}$ with each bound eigenstate. Each of these states has $\langle \hat{n}_{\mathrm{init},\pm}^{(3)}\rangle=2+1/\sqrt{2}$ atoms on the three initial sites [Fig.~\ref{fig_Triplets}(c) inset]. We then find that a 3-tuplet should have $\langle \hat{n}_{\mathrm{init}}^{(3)}\rangle \approx 2.2$ atoms localized, which matches direct numerics in Fig.~\ref{fig_Triplets}(c).

This self-localization shows a stark difference between open and periodic boundary conditions, as evident from the density of an initial 3-tuplet over longer timescales [Figs.~\ref{fig_Triplets}(d-e)]. For periodic boundaries, the 3-tuplet position eventually dissolves, as a 2-tuplet can wrap around the lattice and collide with the orphan from the opposite side. If another 2-tuplet then shoots off, now involving the old orphan, the new orphan will have moved outside the initial three sites. The timescale for this dissolving process grows with system size, since 2-tuplets must cross the entire lattice. For open boundaries, the 2-tuplets can only rebound off the wall and come back, allowing localization to persist. This is indicative of the presence of quantum few-body scars~\cite{Heller}, as the system exhibits closed trajectories with reduced dimension of accessible Hilbert space size, changing the long-time behaviour.

\begin{figure}
\centering
\includegraphics[width=1\linewidth]{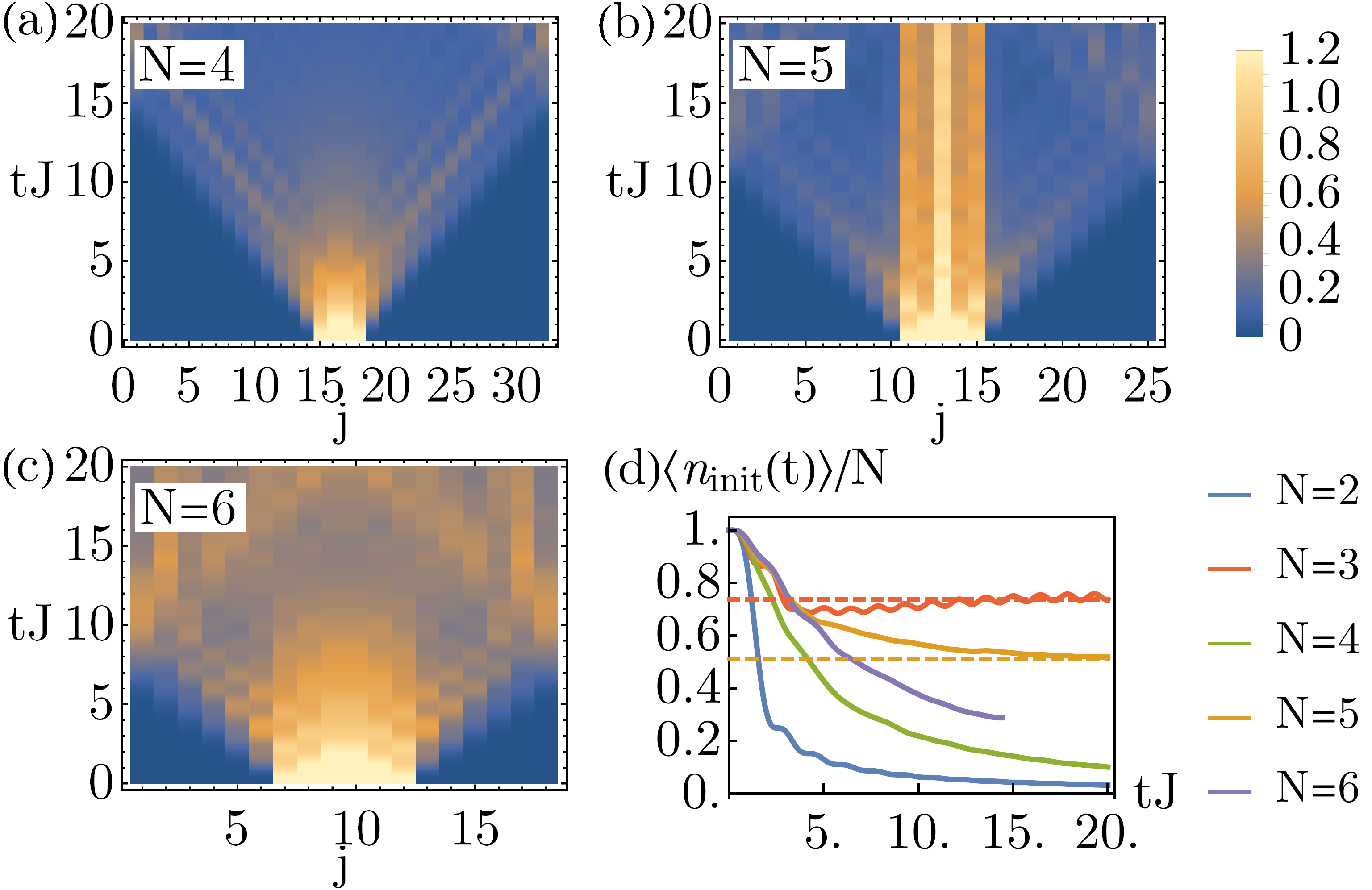}
\caption{Density dynamics of an $N$-tuplet of $\uparrow$ atoms, for $N=4$ (a), $N=5$ (b) and $N=6$ (c). Localization persists for odd $N=5$, but not the even cases. (d) Total number of remaining atoms in the initially-populated sites, for $N=2,3,4,5,6$. System sizes are $L=36,35,32,25,18$. The red dashed line is the analytic prediction for $N=3$ (c.f. Appendix~\ref{app_TripletBoundStates}). The orange dashed line is a quasi-numeric prediction from a bound-state search for $N=5$ (c.f. Appendix~\ref{app_5TupletLocalization}). The even lines are expected to stabilize at mean density $N/L$ in the long-time limit. Note that the $N=6$ line is stopped early to avoid rebounds from the edges.}
\label{fig_NTuplets}
\end{figure}

The multi-body self-localization shown in Fig.~\ref{fig_Triplets} can be interpreted as a coherent quantum version of entropic confinement, where the system tends to stay in states with higher connectivity $\nu$ to reduce its kinetic energy. This is analogous to the three-body bound-states described by Efimov physics, although the system itself is quite different. While the higher energy scales in the system ($U=\Omega$) are not present in the effective Hamiltonian, they manifest indirectly by energetically enforcing restrictions on the motion. Localization is also present in higher-order $N$-tuplet structures ($N$ neighbouring $\uparrow$ singlons), as seen in Fig.~\ref{fig_NTuplets}. We find that even-$N$ configurations will smear out quickly, while odd-$N$ configurations exhibit long-time localization for similar reasons (c.f. Appendix~\ref{app_5TupletLocalization}).

\textit{Scattering.}
We can use the intuitions developed above to understand the 2-tuplet and orphan scattering in Figs.~\ref{fig_DoubletMotion}(c-d). As discussed in the previous section and Appendix~\ref{app_TripletBoundStates}, the Hamiltonian for 3-tuplets can be reduced to a tight-binding chain with a single impurity state shifted in energy (though there are two stuck states, one linear combination decouples). We can further remodel the scattering as a single-particle problem, with a quasiparticle (corresponding to a 2-tuplet) moving though a kronecker-delta potential formed by the impurity (corresponding to an orphan). Under this picture, Eq.~\eqref{eq_AndersonModel} can be rewritten as $\hat{H}^{(3)}_{\mathrm{eff}}=-J\sum_{\langle m,n\rangle}\left(\hat{d}_{m}^{\dagger}\hat{d}_{n}+h.c.\right)-U_{\delta}\hat{d}_{0}^{\dagger}\hat{d}_{0}$ for a potential depth $U_{\delta}$. We estimate the transmission by initializing a particle at $m_{\mathrm{init}} < 0$ and evaluating its density at $m>0$ for times $tJ \gtrsim |m_{\mathrm{init}}|$, rescaling the time and amplitude by 2 (since a 2-tuplet takes two steps to move one real lattice site, and has two atoms).

Fig.~\ref{fig_DoubletMotion}(d) compares this estimated transmission amplitude with direct three-atom numerics. With a potential depth of $U_{\delta} = 2J$, corresponding to the same bound-state localization length $b$ in the single- and three-particle problems, we find good agreement. This simplification also lets us find an analytic expression for the transmission, yielding $\langle \hat{n}_{\mathrm{trans}}\rangle=1-1/\sqrt{2} \approx 0.29$ (c.f. Appendix~\ref{app_TransmissionCoefficient}). The agreement in Fig.~\ref{fig_DoubletMotion}(d) showcases the reduction of a three-body interacting problem to that of a single particle with tractable emergent interactions.

The nontrivial feature not captured by this mapping is the non-reciprocal position dependence of the scatterer. Consider an orphan initially on the right of the three sites of an imminent collision. If the 2-tuplet reflects, then the orphan remains in place.  If transmission occurs, then a new 2-tuplet shoots off to the right, leaving a new orphan two sites left of the original orphan's location. The system exhibits non-reciprocal behaviour where the scatterer is shifted by two sites upon transmission, but not upon reflection. In a classical description, its effective mass is infinite for reflection but finite (and set by quantized motion) for transmission. This behavior is evident in the vertical stripe of atomic density two sites left of the orphan in Fig.~\ref{fig_DoubletMotion}(c). Finding such non-reciprocal effects in a purely closed system as a consequence of interaction-induced constraints is an intriguing feature impossible to capture by the Anderson mapping in Eq.~\eqref{eq_AndersonModel}.

\textit{Experimental implementations.}
Realizing this system is straightforward with a 3D optical lattice populated by two-level atoms. A suggested implementation with e.g.~${}^{87}$Sr is to set the confinement to $V_{x} \sim 8$--$20 E_R$, and $V_{y},V_{z} \gtrsim 100 E_R$ (with $E_R$ the lattice recoil energy), such that on-site repulsion satisfies $U/J \sim 30$--$500$~\cite{zhang2014spectroscopic}. The dynamical timescale will then be set by $J \sim 10$--$100$ Hz. To generate the desired SOC one could use an optical transition in alkaline-earth atoms, or a Raman transitions in alkali atoms.
In both cases, it is required that the laser-imparted phase along $\hat{x}$, is $\pi$ per lattice site. A coherence time much longer than the one set by $J$ is also required. This condition is much more favorable that the stringent condition required to observe superexchange interactions (set by $J^2/U$).

Measurement and preparation of the few-body atomic structures are easiest to do with quantum gas microscopes and optical tweezers respectively. However, a state-of-the-art optical lattice clock such as the $^{87}$Sr clock~\cite{campbell2017strontiumClock,wall2016strontiumSOC} can also probe the 3-tuplet localization without single-site resolution. A sparse configuration of many 3-tuplets can be generated with a lattice tilt and a pulse sequence capable of spectroscopically resolving a specific set of transitions (c.f. Appendix~\ref{app_TripletPreparation}). Allowing the system to evolve and then undoing the pulse sequence to detect the percentage of the atoms remaining in the initial triply-occupied sites can be used to verify localization.

\textit{Conclusions and outlook.}
We have proposed a simple and intuitive system where spin-orbit coupling and interactions generate constrained dynamics. This system exhibits features of self-localization and non-reciprocity, and maps to non-interacting models while maintaining nontrivial lattice effects. A vast number of extensions can be considered, especially if one looks to higher densities or higher dimensions. At higher densities, $\downarrow$ singlons will act as bottlenecks, generically producing many-body quantum scars. Even with no $\downarrow$ singlons, a transfer matrix calculation analogous to~\cite{KN} reveals that in the thermodynamic limit there will be an exponentially large subspace of $2^L$ product states which will be eigenstates of the dynamics, and hence perfect many-body quantum scars (c.f. Appendix~\ref{app_Scars}). The system is also simple to generalize to higher dimensions, needing only an SOC drive with a $\pi$ phase along all allowed directions. The underlying physics can be connected to dynamical gauge fields, where atomic motion by one species is influenced by another in a non-reciprocal manner~\cite{banerjee2012gaugeFields,barbiero2018gaugeFields,schweizer2019gaugeFields}. Higher filling fractions lead to long-range doublon correlations, which can be relevant to studies of superconductivity. Fermionic statistics will also play a significant role in higher dimensions and interplay with the scar and non-reciprocity features. The accessible implementation and the rich breadth of physics displayed make the proposed setup an ideal playground for future investigations.

\textit{Acknowledgements}
M.M. acknowledges support from a CTQM graduate fellowship.
I.K. was supported by a National Research Council Fellowship through the National Institute of Standards and Technology.
R.M.N. was supported by the AFOSR grant FA9550-17-1-0183.
A.M.R is supported by the AFOSR grant FA9550-18-1-0319 and its MURI Initiative, by the DARPA and ARO grant W911NF-16-1-0576, the DARPA DRINQs grant, the ARO single investigator award W911NF-19-1-0210,  the NSF PHY1820885, NSF JILA-PFC PHY1734006 grants, and by NIST.

\bibliographystyle{unsrt}
\bibliography{LeapFrogBibliography.bib}

\clearpage

\onecolumngrid
\appendix

\section{3-Tuplet bound eigenstates}
\label{app_TripletBoundStates}
In the main text, we discussed how to map the three-atom problem to an effective tight-binding chain with two extra stuck states on the central site. We re-emphasize this mapping here. The Hilbert space can be written in terms of outward-moving 2-tuplet states $\ket{s_{j,\alpha}}$ ($\ket{d_{j,\alpha}}$), where $s$ ($d$) is two $\uparrow$ singlons (one doublon), $j$ is the distance of the closest atom from the center of the initial 3-tuplet, and $\alpha \in \{l,r\}$ tells whether the 2-tuplet is left or right of the center. Sample Fock states for the left-side $\uparrow$-singlon pair are (with the center of the triplet underlined),
\begin{equation}
\begin{aligned}
\ket{s_{1,l}}&=\ket{\dots 0,0,0,0,\uparrow,\uparrow,\underline{0},\uparrow,0,0,\dots},\\
\ket{s_{2,l}}&=\ket{\dots 0,0,0,\uparrow,\uparrow,0,\underline{0},\uparrow,0,0,\dots},\\
\ket{s_{3,l}}&=\ket{\dots 0,0,\uparrow,\uparrow,0,0,\underline{0},\uparrow,0,0,\dots},
\end{aligned}
\end{equation}
the right-side singlon pair are,
\begin{equation}
\begin{aligned}
\ket{s_{1,r}}&=\ket{\dots 0,0,\uparrow,\underline{0},\uparrow,\uparrow,0,0,0,0,\dots},\\
\ket{s_{2,r}}&=\ket{\dots 0,0,\uparrow,\underline{0},0,\uparrow,\uparrow,0,0,0,\dots},\\
\ket{s_{3,r}}&=\ket{\dots 0,0,\uparrow,\underline{0},0,0,\uparrow,\uparrow,0,0,\dots},
\end{aligned}
\end{equation}
the left-side doublon are,
\begin{equation}
\begin{aligned}
\ket{d_{1,l}}&=\ket{\dots 0,0,0,0,\uparrow\downarrow,\underline{0},\uparrow,0,0,\dots},\\
\ket{d_{2,l}}&=\ket{\dots 0,0,0,\uparrow\downarrow,0,\underline{0},\uparrow,0,0,\dots},\\
\ket{d_{3,l}}&=\ket{\dots 0,0,\uparrow\downarrow,0,0,\underline{0},\uparrow,0,0,\dots},
\end{aligned}
\end{equation}
and the right-side doublon are,
\begin{equation}
\begin{aligned}
\ket{d_{1,r}}&=\ket{\dots 0,0,\uparrow,\underline{0}\uparrow\downarrow,0,0,0,0,\dots},\\
\ket{d_{2,r}}&=\ket{\dots 0,0,\uparrow,\underline{0},0\uparrow\downarrow,0,0,0,\dots},\\
\ket{d_{3,r}}&=\ket{\dots 0,0,\uparrow,\underline{0},0,0,\uparrow\downarrow,0,0,\dots}.
\end{aligned}
\end{equation}
Aside from these, and assuming open boundaries, there are only three other states in the Hilbert space. These are the 3-tuplet itself,
\begin{equation}
\ket{\uparrow,\uparrow,\uparrow}=\ket{\dots 0,0,0,\uparrow,\underline{\uparrow},\uparrow,0,0,0,\dots},\\
\end{equation}
and the two extra stuck states it can access,
\begin{equation}
\begin{aligned}
\ket{\uparrow,\uparrow\downarrow,0}&=\ket{\dots 0,0,0,\uparrow,\underline{\uparrow\downarrow},0,0,0,0,\dots},\\
\ket{0,\uparrow\downarrow,\uparrow}&=\ket{\dots 0,0,0,0,\underline{\uparrow\downarrow},\uparrow,0,0,0,\dots}.
\end{aligned}
\end{equation}

Since the system tunnels directly along the sequence of states $\ket{d_{j,\alpha}}\to \ket{s_{j,\alpha}}\to \ket{d_{j+1,\alpha}}\to \dots$, we can treat each of those states as its own virtual lattice site, and write down an explicit tight-binding model (as in the main text),
\begin{equation}
\label{eq_app_AndersonHamiltonian}
    \hat{H}^{(3)}=-J\sum_{\langle m,n\rangle}\left(\hat{d}_{m}^{\dagger}\hat{d}_{n}+h.c.\right)-J\left(\hat{d}_{0}^{\dagger}\hat{d}_{L}+\hat{d}_{0}^{\dagger}\hat{d}_{R}+h.c.\right),
\end{equation}
where $\hat{d}_{m}$ annihilates a particle at virtual site $m$, which runs over the corresponding chain of Fock states for the three-body problem (thus two $m$ per $j$ lattice site). The $m=0$ lattice site corresponds to the state $\ket{\uparrow,\uparrow,\uparrow}$, which connects to $\ket{d_{1,l}}$ and $\ket{d_{1,r}}$, and $\hat{d}_{L}$, $\hat{d}_{R}$ annihilation operators correspond to the stuck states $\ket{\uparrow,\uparrow\downarrow,0}$, $\ket{0,\uparrow\downarrow,\uparrow}$.

Eq.~\eqref{eq_app_AndersonHamiltonian} is an explicit interaction-free Anderson model with two impurities. We first rewrite it as a single impurity by rotating the impurity states, since the combination $(\hat{d}_{L}-\hat{d}_{R})/\sqrt{2}$ decouples from the dynamics. Further making a Fourier transform $\hat{d}_{m}=\frac{1}{\sqrt{2L}}\sum_{k}e^{ikm}\hat{d}_{k}$ (ignoring the boundaries), we arrive at,
\begin{equation}
    \hat{H}^{(3)}=-2J\sum_{k}\cos(k)\hat{d}_{k}^{\dagger}\hat{d}_{k}-\frac{\sqrt{2}J}{\sqrt{2L}}\sum_{k}\left(\hat{d}_{k}^{\dagger}\hat{d}_{C}+h.c.\right),
\end{equation}
with $\hat{d}_{C}=(\hat{d}_{L}+\hat{d}_{R})/\sqrt{2}$ corresponding to the coupled impurity state. At this point, we can follow Ref.~\cite{kehrein1996anderson} to obtain an explicit expression for the energy shift $E$ of the impurity state, which yields,
\begin{equation}
    E=\sqrt{2(1+\sqrt{2})}J.
\end{equation}

A tight-binding chain with a single state shifted in energy suggests the presence of an exponentially localized bound-state with the same energy in the full three-atom problem.
By observing that the eigenvectors of $\hat{H}^{(3)}$ must be symmetric under reflection about the origin, we find the bound-states through the use of the ansatz
\begin{align}
\ket{\phi_{\mathrm{ansatz}}} = A \ket{\uparrow,\uparrow,\uparrow}+B (\ket{\uparrow,\uparrow\downarrow,0}+\ket{0,\uparrow\downarrow,\uparrow}) &+D\sum_{j=1}^{L_d+1}b^{-(2j-1)}\left(\ket{d_{j,l}}+\ket{d_{j,r}}\right) \notag \\
&+S\sum_{j=1}^{L_d}b^{-2j}\left(\ket{s_{j,l}}+\ket{s_{j,r}}\right),
\end{align}
where $A,B,D,S,b$ are free parameters, with localized weights $A,B\in\mathbb{R}$, signs $D,S\in\set{\pm1}$, and exponential localization falloff $b\in\mathbb{R}$ (i.e.~the localization length in units of the virtual lattice spacing). Testing this ansatz, we find two matching bound eigenstates,
\begin{equation}
\begin{aligned}
\ket{\phi_{\pm}} =\mathcal{N}\bigg[\pm\ket{\uparrow,\uparrow,\uparrow}+\frac{1}{\sqrt{2}b}\left(\ket{\uparrow,\uparrow\downarrow,0}+\ket{0,\uparrow\downarrow,\uparrow}\right)&+\sum_{j=1}^{L_d+1}(-1)^{j}b^{-(2j-1)}\left(\ket{d_{j,l}}+\ket{d_{j,r}}\right)\\
&\pm \sum_{j=1}^{L_d}(-1)^{j}b^{-2j}\left(\ket{s_{j,l}}+\ket{s_{j,r}}\right)\bigg],
\end{aligned}
\end{equation}
where $b=\sqrt{1+\sqrt{2}}\approx1.6$ is the localization length, $L_d$ is the number of empty sites on either side of the triplet, and $E=\sqrt{2}bJ$ is the same as found earlier in the Anderson model. These eigenstates are exact up to exponentially small corrections $\mathcal{O}\left(b^{-2L_d}\right)$ on the boundaries. The normalization constant depends on system size, but equals $\mathcal{N} = 1/\sqrt{2\sqrt{2}}$ in the $L_d \to \infty$ limit.

Assuming sufficient padding (i.e. $L_d \to \infty$), we can immediately compute overlaps and expectation values. An initially-localized 3-tuplet has bound-state overlap,
\begin{equation}
\lim_{L_d \to \infty}\left|\bra{\uparrow,\uparrow,\uparrow}\phi_{\pm}\rangle\right|^2 = \frac{1}{2\sqrt{2}} \approx 0.35.
\end{equation}
Each bound eigenstate has a density profile as given in main text Fig.~\ref{fig_Triplets}(c) inset. For these eigenstates, the middle three sites have a localized population of
\begin{equation}
\lim_{L_d \to \infty}\langle \hat{n}_{\mathrm{init,\pm}}\rangle = 2 + \frac{1}{\sqrt{2}} \approx 2.71 \>\>\> \text{for } \ket{\phi_{\pm}}.
\end{equation}
From these, we can obtain the localized population of a 3-tuplet,
\begin{equation}
\label{eq_TripletLocalizedPopulation}
\begin{aligned}
  \langle \hat{n}_{\mathrm{init}}^{(3)}\rangle&|_{t\to \infty} = \sum_{\pm}|\langle \uparrow,\uparrow,\uparrow|\phi_{\pm}^{(3)}\rangle|^2\times\langle \hat{n}_{\mathrm{init},\pm}^{(3)}\rangle \\
  &+\left(1-\sum_{\pm}|\langle \uparrow,\uparrow,\uparrow|\phi_{\pm}^{(3)}\rangle|^2\right)\times 1=\frac{3}{2}+\frac{1}{\sqrt{2}}\approx 2.2.
\end{aligned}
\end{equation}
This is the value predicted in the main text. Note that the term on the second line is the contribution from any state population \textit{not} in the bound-states, which will still have at least one atom remain as an orphan assuming open boundaries.

\section{2-Tuplet and orphan transmission coefficient}
\label{app_TransmissionCoefficient}
In the maintext, we showed that the number of atoms transmitted from a 2-tuplet and orphan collision could be approximated by a single particle propagating past a Kronecker delta potential of strength $U_{\delta} = 2J$, in line with the Anderson model mapping. The corresponding Hamiltonian is $\hat{H}^{(3)}_{\mathrm{eff}}=-J\sum_{\langle m,n\rangle}\left(\hat{d}_{m}^{\dagger}\hat{d}_{n}+h.c.\right)-U_{\delta}\hat{d}_{0}^{\dagger}\hat{d}_{0}$. The transmission coefficient is given by $\langle \hat{n}_{\mathrm{trans}}=\rangle\sum_{m > 0} \langle \hat{n}_{m}\rangle$ with $m=0$ the initial position of the orphan (i.e. the scatterer), assuming that the 2-tuplet starts at $m_\mathrm{init} < 0$ with at least a few sites of space between it and the orphan. The long-time transmission is obtained by evaluating $\langle \hat{n}_{\mathrm{trans}}\rangle$ at time $tJ > |m_{\mathrm{init}}|$, i.e. long after the collision has finished and assuming sufficient padding on both sides to avoid rebound effects.

We can evaluate this long-time coefficient analytically with a Green's function calculation. Following Ref.~\cite{economou1983green}, the transmission coefficient for a 1D tight-binding chain with a single site shifted in energy by $U_{\delta}$ is given by,
\begin{equation}
    t(\epsilon) = \frac{1}{\left|1- U_{\delta}G_0(0,0;\epsilon)\right|^2},
\end{equation}
where $\epsilon$ is the energy of the state being scattered, and $G_0 (0,0;\epsilon)$ is the diagonal element of the unperturbed (i.e. 1D tight-binding chain) lattice Green's function at the energy-shifted site ($m=0$). This Green's function may be written as~\cite{economou1983green},
\begin{equation}
    G_0(l,l;\epsilon)=\frac{-i}{\sqrt{4J^2 - \epsilon^2}}.
\end{equation}
Since the spectrum of the tight-binding chain is the simple cosine dispersion of $\epsilon = -2J \cos (k)$, and we already know $U_{\delta} = 2J$, we can rewrite the transmission coefficient as a function of lattice quasimomentum,
\begin{equation}
    t(k) = \frac{1}{1+\frac{1}{1-\cos(k)^2}}.
\end{equation}
Our particular scattering problem uses a localized wavepacket (i.e. a single atom at $j_0$), which is an equal superposition of all quasimomentum modes. We can thus write the total transmission coefficient as an equal-weight average of $t(k)$ for all allowed $k \in [-\pi,\pi]$. Note that the result we get will have to be both divided by two, because the quasimomentum modes have no notion of direction and we are looking only at population ending up to the right of the scatterer, and multiplied by two because the quasiparticle has two atoms. Altogether, the total transmission coefficient becomes,
\begin{equation}
    \langle \hat{n}_{\mathrm{trans}}\rangle = 2 \times \frac{1}{2}\times \frac{1}{2\pi}\int_{0}^{2\pi}t(k) dk = 1-\frac{1}{\sqrt{2}} \approx 0.29.
\end{equation}
This value matches the long-time results in main text Fig.2(d), and tells us how many atoms of a 2-tuplet get past an orphan during their collision.

\section{5-Tuplet numerical localization}
\label{app_5TupletLocalization}
In this appendix, we examine the case of $5$-tuplet dynamics (five initially neighbouring $\uparrow$ atoms) to generalize our bound-state results. While the 3-tuplet case could be solved analytically since it mapped to an effective 1D tight-binding chain, the 5-tuplet would instead span a 2D space because it consists of up to two travelling 2-tuplets, rendering analytics difficult. We instead turn to numerics.

The relevant eigenspectrum of the system is obtained by first running a full simulation of Fermi-Hubbard dynamics with the entire Fock space to a sufficiently long time $tJ \gg 1$, from an initial condition where the 5-tuplet is at the center of the lattice (open boundaries, $L=19$). All states with nonzero resulting population to machine precision are kept as the accessible states for the 5-tuplet, which reduces the size of the Hilbert space from the full binomial ${2L}\choose{5}$ to a few hundred states. Rewriting the Hamiltonian in the basis of these states allows an explicit search for bound-states from their respective population profiles.

\begin{figure}
\centering
\includegraphics[width=0.7\linewidth]{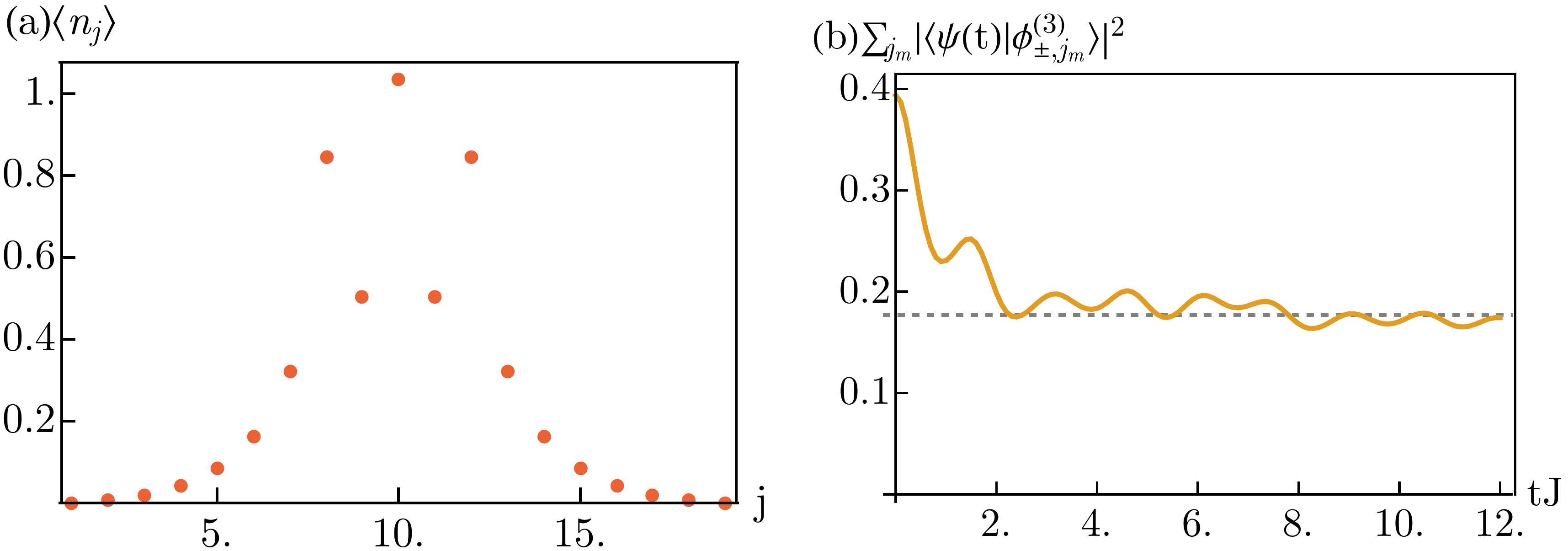}
\caption{(a) Density profile of the 5-body bound eigenstates $\ket{\phi_{\pm}^{(5)}}$ of the 5-tuplet system. (b) Total overlap of the 5-tuplet wavefunction with all possible states $\ket{\phi_{\pm,j_m}^{(3)}}$ of an outgoing 2-tuplet and 3-body bound eigenstate (on the left side of the 5-tuplet, without loss of generality) as described in the main text, summing over both such states for all possible $j_m$.}
\label{fig_5BodyTripletOverlap}
\end{figure}

We find two full 5-body bound-states $\ket{\phi_{\pm}^{(5)}}$, whose density profiles are depicted in Fig.~\ref{fig_5BodyTripletOverlap}(a) (equal for both). Due to the exponential tails, the total amount of atoms present in the central 5 sites for each eigenstate is $\langle \hat{n}_{\mathrm{init},\pm}^{(5)}\rangle \approx 3.71$. The overlap between each of these states and the 5-tuplet is $|\langle \uparrow,\uparrow,\uparrow,\uparrow,\uparrow| \phi_{\pm}^{(5)}\rangle|^2 \approx 0.062$. The contribution to localized atom number from these states alone is not enough to describe the amount we see. The rest comes from effective 3-body states that can also be present in the problem, after a 2-tuplet has been shot off.

Finding 3-body bound-states from the spectrum is more difficult, because they will have an outward-bound 2-tuplet whose eigenmodes will be sinusoidal waves (as it bounces back and forth). We approximate the populations of these states by constructing them artificially. We take a 3-body bound-state in a 3-atom Fock space, $\ket{\phi_{\pm}^{(3)}}$, centered on the left three sites of the 5-tuplet's position without loss of generality (corresponding to a 2-tuplet shot to the right). We then augment the Fock space by adding either a doublon on site $j_m$ to the right of the bound state, or a pair of $\uparrow$ singlons on sites $j_m, j_m+1$. Finally, we rewrite the state in this new 5-atom Fock space, keeping the same phases and amplitudes as the 3-body bound-state had. If a given Fock state already had atoms on the sites where we tried to create more, its corresponding amplitude is set to zero in the new (larger) Hilbert space. Effectively, we have thus constructed a bound-state with an outgoing 2-tuplet $\hat{a}_{j_m,\uparrow}^{\dagger}\hat{a}_{j_m,\downarrow}^{\dagger}\ket{\phi_{\pm}^{(3)}}$, or $\hat{a}_{j_m,\uparrow}^{\dagger}\hat{a}_{j_m+1,\uparrow}^{\dagger}\ket{\phi_{\pm}^{(3)}}$ (depending on the 2-tuplet's current form). We next time-evolve a 5-tuplet using full Fermi-Hubbard dynamics, and compute the time-dependent overlaps of the wavefunction with all of the states we have constructed. Summing over all such overlaps will tell us the overlap with a bound three-body eigenstate on the left three sites, and an outgoing 2-tuplet to the right (in all possible eigenmodes). Fig.~\ref{fig_5BodyTripletOverlap}(b) shows this population (equal for $\pm$ bound-states).

The resulting overlap is roughly $\sim 0.17$ per state. Since there are two of these on the left three sites ($\pm$), and another two on the right three sites (for a left-bound 2-tuplet instead), the total overlap with three-body bound-states is found to be $\sim 0.7 \approx 1/\sqrt{2}$, which is equal to what we found in the three-atom system. We thus conclude that the 5-tuplet overlap with 3-body bound-states is the same as for the 3-tuplet case, meaning that its contribution to the localized atom number can be included in the same manner. Altogether, we get,
\begin{equation}
\begin{aligned}
    \langle \hat{n}_{\mathrm{init}}^{(5)}\rangle|_{t\to \infty} &= \sum_{\pm}|\langle \uparrow,\uparrow,\uparrow,\uparrow,\uparrow|\phi_{\pm}^{(5)}\rangle|^2\times\langle \hat{n}_{\mathrm{init},\pm}^{(5)}\rangle
    +\sum_{j_m}'\sum_{\pm}|\langle \uparrow,\uparrow,\uparrow,\uparrow,\uparrow|\phi_{\pm,j_m}^{(3)}\rangle|^2\times\langle \hat{n}_{\mathrm{init},\pm}^{(3)}\rangle \\
    &+\left(1-\sum_{\pm}|\langle \uparrow,\uparrow,\uparrow,\uparrow,\uparrow|\phi_{\pm}^{(5)}\rangle|^2-\sum_{j_m}'\sum_{\pm}|\langle \uparrow,\uparrow,\uparrow,\uparrow,\uparrow|\phi_{\pm,j_m}^{(3)}\rangle|^2\right)\times 1\approx 2.54,
\end{aligned}
\end{equation}
where the primed sum over $j_m$ runs over all possible 2-tuplet configurations on all sites $j_m$ to the right of the three where the three-body bound-state is centered (including both doublon and two-singlon configurations). The term on the second line is all non-bound population with one orphan remaining. This estimate is in good agreement with the direct numerics in the main text Fig.~4(d).

\section{3-Tuplet preparation}
\label{app_TripletPreparation}
In order to facilitate experimental investigation of the confinement phenomena discussed in this work, here we sketch an example experimental protocol for preparing isolated 3-tuplets on an effectively 1D lattice.
In particular, we consider the preparation of isolated 3-tuplets in the 3D ${}^{87}$Sr optical lattice clock, where tunneling can be restricted to 1D.
Our protocol does not require single-site addressability, and consists of four stages: (i) initializing a high-density gas of ultracold atoms on a 3D lattice, (ii) removing all atoms except those in the subspace of three nuclear spin states, (iii) removing all atoms except those in triply-occupied lattice sites, and (iv) converting triply-occupied lattice sites into 3-tuplets (i.e.~spin-polarized ground-state atoms on three neighboring lattice sites) on a lattice.

We begin by loading ground-state ultracold ${}^{87}$Sr atoms into a three-dimensional primitive cubic optical lattice at the ``magic'' wavelength ($\approx813$ nm) for which both the ${}^1S_0$ and ${}^3P_0$ electronic states of the atoms (i.e.~the ``clock states'') see the same lattice potential~\cite{ye2008quantum}.
These atoms should be prepared in a uniform mixture of their (ten) nuclear spin states, at an atom density that (ideally) maximizes the number of triply-occupied lattice sites, as demonstrated in Ref.~[\citenum{goban2018emergence}].
We then turn on an external magnetic field that, due to hyperfine coupling, induces nuclear-spin-dependent (Zeeman) shifts to the electronic excitation energies of the atoms.
The single-atom ${}^1S_0$--${}^3P_0$ transition, for example, gets shifted (to first order) by $\sim100\times m_F$ Hz/gauss in the presence of a magnetic field, where $m_F\in\set{-9/2,-7/2,\cdots,9/2}$ is projection of an atom's nuclear spin onto the magnetic field axis~\cite{boyd2007high}.
By addressing nuclear-spin-dependent electronic transitions, we can selectively excite atoms in all but three nuclear spin states (e.g.~$9/2,7/2,5/2$), and subsequently remove these excited atoms from the lattice.
Note that in order for these excitations to be nuclear-spin selective on multiply-occupied lattice sites, the magnetic shifts to the electronic transitions must be large compared to on-site interaction energies (that are on the scale of kHz~\cite{goban2018emergence, perlin2019effective}).
The presence of interaction energy shifts will also require several excite-remove cycles in order to address lattice sites with different numbers of atoms.

Once we have removed all but three nuclear spin states on the lattice, we wish to vacate all singly- and doubly-occupied lattice sites.
Here, we can make use of the inter-atomic interaction energy shifts to electronic excitation energies~\cite{goban2018emergence, perlin2019effective} by selectively exciting, and subsequently removing, only atoms on doubly- and singly-occupied lattice sites.
At this point, the lattice should be mostly vacant, with roughly one in ${10 \choose 3}=120$ lattice sites containing three atoms that occupy the previously chosen nuclear spin states.

\begin{figure}
\centering
\includegraphics[width=0.5\linewidth]{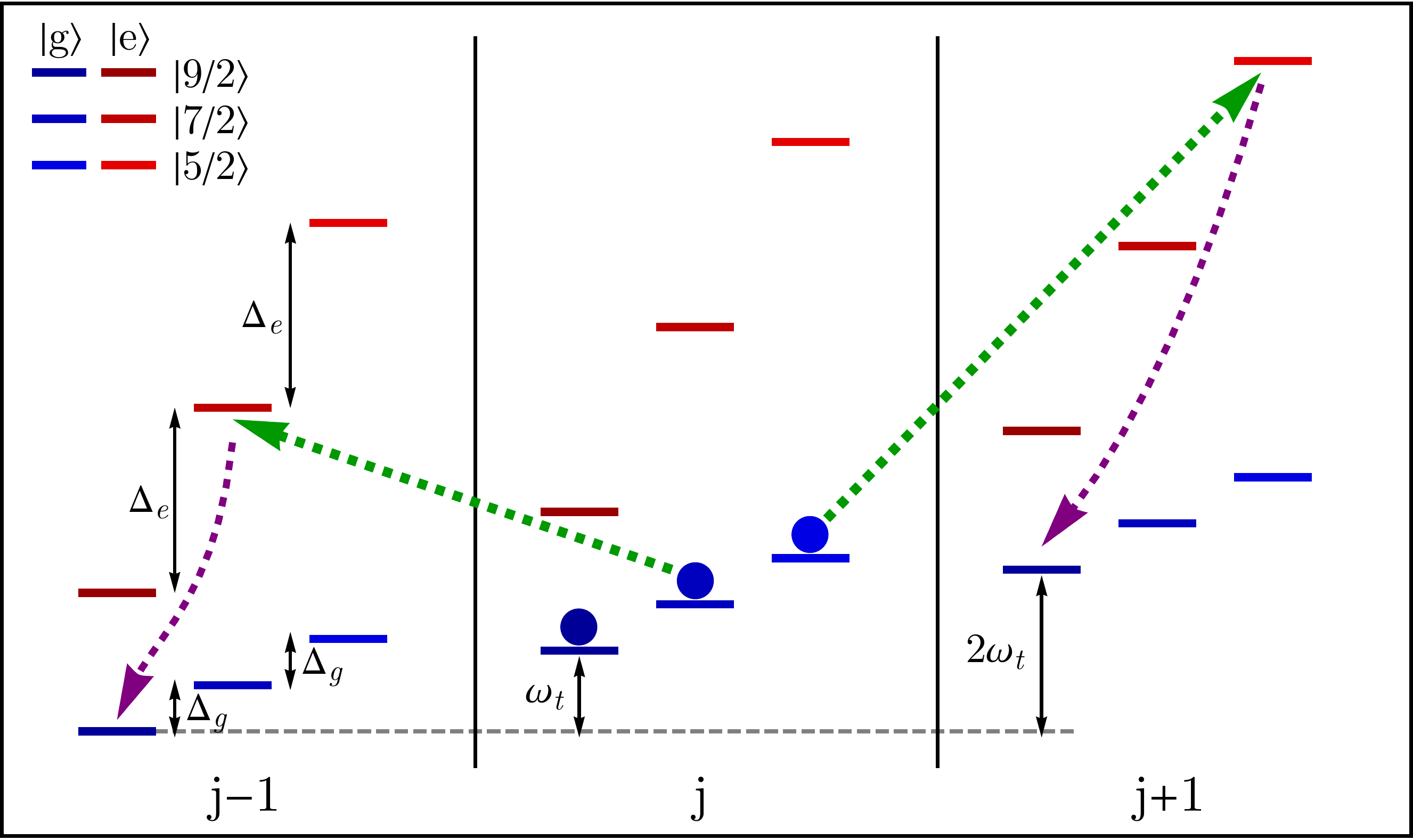}
\caption{Schematic for the preparation of a 3-tuplet from an initially triply-occupied lattice site $j$.  Blue and red lines respectively indicate ground and excited electronic energy levels of individual atomic states, with different shades corresponding to particular nuclear spin states.  An external linear potential induces an energy difference $\omega_t$ between neighboring lattice sites, and a magnetic field induces nuclear-spin-dependent energy shifts $\Delta_g\times m_F$ ($\Delta_e\times m_F$) for atoms in the ground (excited) electronic state, with $m_F$ the nuclear spin of an atomic state.  With both left/right symmetry and nuclear-spin degeneracy broken, an external laser can selectively induce directional tunneling of atoms in particular nuclear spin states by tuning the laser frequency to the desired transition (dashed green arrows).  Once one atom has been moved to each of the lattice sites on either side of $j$, the excited atoms can be brought down to their electronic ground state and pumped into a single nuclear spin (dashed purple arrows), thereby preparing a 3-tuplet.  For simplicity, this figure does not consider the effect of inter-atomic interactions.}
\label{fig_PreparationSchematic}
\end{figure}

Our final task is to convert triply-occupied lattice sites into 3-tuples, which can be done through a protocol schematically shown in Fig.~\ref{fig_PreparationSchematic}.
The basic idea is to (i) apply a linear potential gradient (e.g.~using linear Stark shifts) in order to energetically distinguish atomic states on different lattice sites, and then (ii) additionally turn on a magnetic field to induce nuclear-spin-selective (Zeeman) shifts to the electronic excitation energies of the atoms.
By breaking both left/right symmetry and the nuclear spin degeneracy of electronic excitation energies, we can selectively induce directional tunneling of atoms in particular nuclear spin states.
Specifically, we can tune the frequency of an external laser to resonance on a transition that e.g.~excites an atom with nuclear spin 7/2 and moves it to the left by one lattice site.
Note that interaction energies, coming from both two- and three-body interactions~\cite{goban2018emergence, perlin2019effective}, need to be taken into account when tuning laser frequencies to resonance on a desired transition.
Once one atom has been moved to each of the lattice sites on either side of the central (initially triply-occupied) lattice site, the excited atoms on the sides can be brought down to their electronic ground state and all atoms can be pumped into a single nuclear spin state, thereby preparing a 3-tuplet.

\section{Model robustness}
\label{app_ModelAgreement}
We want to benchmark the agreement between the gauged Fermi-Hubbard model and density-dependent tunneling model [Eqs.~\eqref{eq_FermiHubbardGauged} and~\eqref{eq_FermiHubbardAsymptotic}]. Fidelity may not be a good measure as it can fall off with system size while relevant observables remain in agreement. We instead look at the dynamics of correlators. One of the key features of the model is its specific form of doublon production stemming from the leap-frog mechanism. We look at average doublon number as a function of time for the two models, and define a metric for the average error over the timespan,
\begin{equation}
\label{eq_DoublonError}
\epsilon (\langle \hat{n}_d\rangle) = \left[\frac{1}{t_f}\int_{0}^{t_f} dt \left| \frac{\langle \hat{n}_d\rangle(t)-\langle \hat{n}_{d,\mathrm{id}}\rangle(t)}{\langle \hat{n}_{d,\mathrm{id}}\rangle(t)}\right|^2\right]^{1/2},
\end{equation}
where $\hat{n}_{d} = \sum_{j} \hat{n}_{j,\uparrow}\hat{n}_{j,\downarrow}/L$ is average doublon number, the expectation $\langle \hat{n}_{d}\rangle$ is evaluated from time-evolution of the full Fermi-Hubbard in Eq.~\eqref{eq_FermiHubbardGauged}, while $\langle \hat{n}_{d,\mathrm{id}}\rangle$ uses the ideal model of Eq.~\eqref{eq_FermiHubbardAsymptotic}. In a sense, $\epsilon (\langle \hat{n}_{d} \rangle)$ acts as a \% error estimate for the dynamics of the correlator. Of course, this metric is imperfect, as it will inevitably become worse over longer times, and assumes that errors in the behaviour of $\langle \hat{n}_{d}\rangle$ accurately capture agreement between the models. We will only use it for a qualitative estimate to show some degree of robustness.

\begin{figure}
\centering
\includegraphics[width=0.85\linewidth]{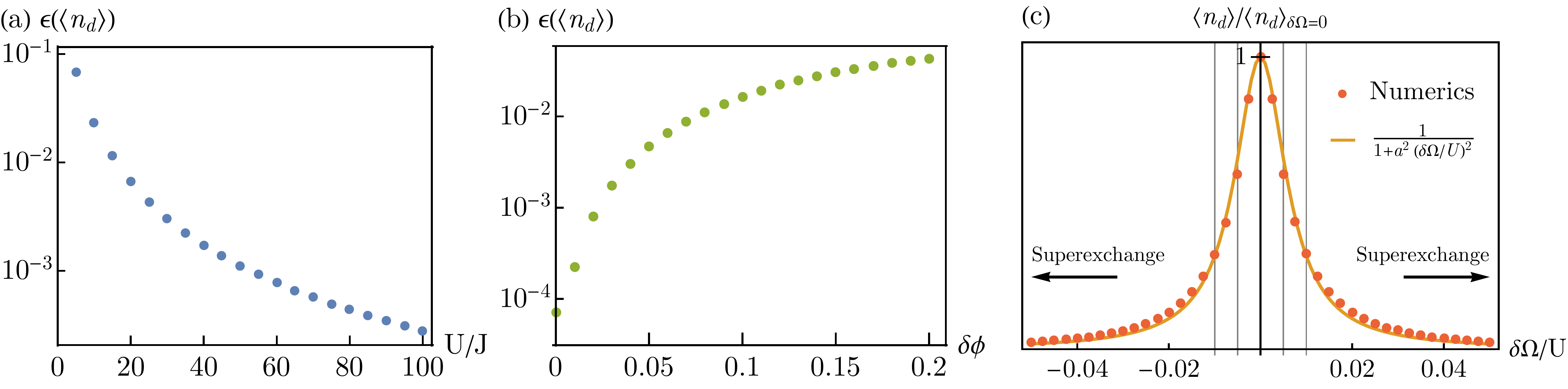}
\caption{(a) Doublon error [Eq.~\eqref{eq_DoublonError}] for increasing $U/J$, using times $t=0$ to $t_f J=10$. Initial loadout is a sample representative configuration: system size of $L=8$ with periodic boundary conditions and $N=6$ atoms, starting in all $\uparrow$-singlons except sites $j=1,4$ which are vacancies. (b) Doublon error for imperfect flux $\delta \phi$ using the same parameters as above, except fixing $U/J = \Omega/J = 200$. (c) Model agreement for $\delta \Omega = U-\Omega \neq 0$. The error metric in prior panels cannot be used due to drastically differing timescales. Mean steady-state doublon number is instead plotted, normalized by the mean number of $\delta \Omega = 0$. We compute $\langle \hat{n}_{d}\rangle$ for $tJ = 0 $ to $50$, and average over $tJ \in [10,50]$ (to avoid major fluctuations at short times). Lattice size and initial filling are the same as prior panels. Interaction strength is $U/J = 200$. Gray vertical lines correspond to intervals of $J$. The orange line is a Lorentzian fit, with coefficient $a \approx 0.73 U/J$. Arrows indicate regimes where $|\delta \Omega| \gg J$, where conventional second-order superexchange holds.}
\label{fig_ModelAgreement}
\end{figure}

The first comparison we make is low $U/J = \Omega/J$. The mapping is exact for $U/J \to \infty$. In Fig.~\ref{fig_ModelAgreement}(a), we plot the error for a specific initial loadout and time interval. The model should be valid provided that $tJ \ll U/J$. Given that it is straightforward to generate $U/J \sim 100$ without significantly compromising experimental timescales, errors rising from this should not be significant.

The next thing to compare is imperfect flux. Recall that in writing the original, ungauged Fermi-Hubbard model, we had an $e^{i j \phi} = (-1)^j$ phase in the drive rising from an effective flux $\phi = \pi$. If we assume that this flux can deviate, $\phi = \pi + \delta\phi$, we can still make a gauge transformation to turn the drive into a $\hat{\sigma}^{z}$ term via,
\begin{equation}
\hat{a}_{j,\uparrow}=\left[\hat{c}_{j,e}+e^{i j \delta \phi}(-1)^j \hat{c}_{j,g}\right]/\sqrt{2},\>\>\>\>\>\hat{a}_{j,\uparrow}=\left[\hat{c}_{j,e}-e^{i j \delta \phi}(-1)^j \hat{c}_{j,g}\right]/\sqrt{2}.
\end{equation}
However, the tunneling will be modified, yielding,
\begin{equation}
\begin{aligned}
\hat{H}_{\delta \phi} = &\frac{J}{2} \sum_{\langle j,k\rangle}\left[\left(1-e^{-i \delta \phi}\right)\left(\hat{a}_{j,\uparrow}^{\dagger}\hat{a}_{k,\uparrow}+\hat{a}_{j,\downarrow}^{\dagger}\hat{a}_{k,\downarrow}\right)+\left(1+e^{-i \delta \phi}\right)\left(\hat{a}_{j,\uparrow}^{\dagger}\hat{a}_{k,\downarrow}+\hat{a}_{j,\downarrow}^{\dagger}\hat{a}_{k,\uparrow}\right)+h.c.\right]\\
&+U \sum_{j}\hat{n}_{j,\uparrow}\hat{n}_{j,\downarrow}+\frac{\Omega}{2}\sum_{j}\left(\hat{n}_{j,\downarrow}-\hat{n}_{j,\uparrow}\right),
\end{aligned}
\end{equation}
which reduces to the gauged Fermi-Hubbard model for $\delta \phi = 0$ as expected. Essentially, erroneous terms arise that let atoms tunnel without flipping their spin (thus breaking the exclusivity of the resonant tunneling). Fig.~\ref{fig_ModelAgreement}(b) shows the error metric in terms of $\delta \phi$, giving error tolerance on the order of $\delta \phi \lesssim 0.1$, which is reasonable assuming the laser can be aimed with angular deviations not exceeding a few degrees.

Finally, we look at the model's discrepancy when the driving is not perfectly matched, i.e. for $\delta \Omega = U -\Omega \neq 0$. In this case, the error metric described above is no longer accurate because the timescales change quickly. An imperfect resonance will slow down the dynamics, although for small deviations they will still give the expected behaviour (just on a different timescale). What we do instead is look at mean doublon number over a long time, as shown in Fig.~\ref{fig_ModelAgreement}(c). When $\delta \Omega/J \lesssim 1$, the doublon number remains high, meaning that the resonance-assisted dynamics are still working. Deviating further causes the doublon number to fall off as a Lorentzian. This drop corresponds to entering the superexchange regime where conventional second-order processes dominate, and the respective perturbation theory denominator proportional to $\delta \Omega$ acts as a good expansion parameter. Tuning $\Omega$ to be within $J$ of $U$ can be made easier by increasing transverse ($\hat{y}$, $\hat{z}$) confinement, which allows a larger $J$ along $\hat{x}$ for the same ratio $U/J$ (and thus more room for error).

\section{Perfect many body scars}
\label{app_Scars}
In this Appendix we sketch a calculation showing the existence of perfect many body scars in the subspace with no $\downarrow$ singlons. (Recall that each $\downarrow$ singlon will itself act as a bottleneck cutting the chain, thereby producing quantum scars in an obvious manner). The calculation follows an inductive argument outlined in Ref.\cite{KN}.

In a system of size $L=2$, in the subspace with no $\downarrow$ singlons, there are six product states that are eigenstates of our Hamiltonian - $\ket{\uparrow,0}$, $\ket{0,\uparrow}$, $\ket{0,0}$, $\ket{\uparrow,\uparrow\downarrow}$, $\ket{\uparrow\downarrow,\uparrow}$ and $\ket{\uparrow\downarrow,\uparrow\downarrow}$. Now suppose $N_r(L)$ is the number of product states with the final site being $r$, which are also eigenstates of the Hamiltonian for a system of size $L$. Adding an additional site will then give another product state that is an eigenstate of the Hamiltonian if and only if the state at site $r$ and the state of the added site together form one of the six trivial states of the $L=2$ Hamiltonian. One may then write down a recursion relation for the number of product states that are also eigenstates of the Hamiltonian, which takes the form
\begin{equation}
\left( \begin{array}{c} N_{\uparrow} \\ N_{0} \\ N_{d} \end{array}\right)_{L+1} = \left(\begin{array}{ccc}  0 & 1 & 1 \\ 1 & 1 & 0 \\ 1 & 0 & 1 \end{array} \right) \left(\begin{array}{c} N_{\uparrow} \\ N_{0} \\ N_{d}\end{array} \right)_L,
\end{equation}
where $N_{\uparrow}$, $N_{0}$ and $N_{d}$ are the numbers of trivial states with their last-added site in the state $\ket{\uparrow}$, $\ket{0}$ and $\ket{\uparrow\downarrow}$ respectively. Given the initial condition $(N_{\uparrow}, N_{0},N_{d})_2 = (2,2,2)$ and the above recursion relation, one may then straightforwardly determine the number of product states that are also eigenstates of the Hamiltonian, by acting repeatedly with the $3\times 3$ transfer matrix above. Since the largest eigenvalue of this matrix is $2$ it then follows that for large system sizes $L$ there are $\sim2^L$ eigenstates of the dynamics which are {\it product} states (zero entanglement) in the natural basis, and which hence constitute an exponentially large set of perfect quantum many body scars.

\end{document}